\documentclass[conference]{IEEEtran}
\IEEEoverridecommandlockouts
\usepackage{cite}
\usepackage{amsmath,amssymb,amsfonts}
\usepackage{hyperref}
\usepackage[ruled, linesnumbered]{algorithm2e}
\SetKwComment{Comment}{/* }{ */}

\usepackage{graphicx}
\usepackage{textcomp}
\usepackage{physics}
\usepackage[inkscapeformat=pdf]{svg}
\usepackage{xcolor}
\def\BibTeX{{\rm B\kern-.05em{\sc i\kern-.025em b}\kern-.08em
    T\kern-.1667em\lower.7ex\hbox{E}\kern-.125emX}}
\begin{document}

\title{Quantum Speedup for the Maximum Cut Problem*\\
\thanks{This work is supported by the National Science Foundation of the Republic of China under MOST 105-2221-E-151-040. R. Wong was supported by the National Science and Technology Council, the Ministry of Education (Higher Education Sprout Project NTU-111L104022), and the National Center for Theoretical Sciences of Taiwan. \\
* Corresponding authors: WL Chang (changwl@cc.kuas.edu.tw), R Wong (renata.wong@phys.ncts.ntu.edu.tw)}
}
\author{\IEEEauthorblockN{1\textsuperscript{st}* Weng-Long Chang}
\IEEEauthorblockA{\textit{Department of Computer Science and Information Engineering} \\
\textit{National Kaohsiung University of Science and Technology}\\
Kaohsiung, Taiwan \\
changwl@cc.kuas.edu.tw}
\and
\IEEEauthorblockN{2\textsuperscript{nd}* Renata Wong}
\IEEEauthorblockA{\textit{Physics Division} \\
\textit{National Center for Theoretical Sciences}\\
\textit{National Taiwan University}\\
Taipei, Taiwan \\
https://orcid.org/0000-0001-5468-0716}
\and
\IEEEauthorblockN{3\textsuperscript{rd} Wen-Yu Chung}
\IEEEauthorblockA{\textit{Department of Computer Science and Information Engineering} \\
\textit{National Kaohsiung University of Science and Technology}\\
Kaohsiung, Taiwan \\
wychung@nkust.edu.tw}
\and
\IEEEauthorblockN{4\textsuperscript{th} Yu-Hao Chen}
\IEEEauthorblockA{\textit{Department of Physics} \\
\textit{National Taiwan University}\\
Taipei, Taiwan \\
r10222067@ntu.edu.tw}
\and
\IEEEauthorblockN{5\textsuperscript{th} Ju-Chin Chen}
\IEEEauthorblockA{\textit{Department of Computer Science and Information Engineering} \\
\textit{National Kaohsiung University of Science and Technology}\\
Kaohsiung, Taiwan \\
jc.chen@nkust.edu.tw}
\and
\IEEEauthorblockN{6\textsuperscript{th} Athanasios V. Vasilakos}
\IEEEauthorblockA{\textit{Center for AI Research (CAIR)} \\
\textit{University of Agder}\\
Grimstad, Norway \\
thanos.vasilakos@uia.no}
}

\maketitle

\begin{abstract}
Given an undirected, unweighted graph with $n$ vertices and $m$ edges, the maximum cut problem is to find a partition of the $n$ vertices into disjoint subsets $V_1$ and $V_2$ such that the number of edges between them is as large as possible. Classically, it is an NP-complete problem, which has potential applications ranging from circuit layout design, statistical physics, computer vision, machine learning and network science to clustering. In this paper, we propose a quantum algorithm to solve the maximum cut problem for any graph $G$ with a quadratic speedup over its classical counterparts, where the temporal and spatial complexities are reduced to, respectively, $O(\sqrt{2^n/r})$ and $O(m^2)$. With respect to oracle-related quantum algorithms for NP-complete problems, we identify our algorithm as optimal. Furthermore, to justify the feasibility of the proposed algorithm, we successfully solve a typical maximum cut problem for a graph with three vertices and two edges by carrying out experiments on IBM’s quantum computer.
\end{abstract}

\begin{IEEEkeywords}
data structures and algorithms, the maximum cut problem, quantum algorithms, quantum computing, quantum speedup
\end{IEEEkeywords}

\section{Introduction}
Let $G = (V, E)$ be an undirected, unweighted graph with a set of vertices $V$ is a set of edges $E$. Further, let $|V|=n$ and let $|E|=m$. A cut $\{V_1, V_2\}$ of $G$ is defined as a partition of vertices into two disjoint subsets $V_1$ and $V_2$. The size of the cut is the number of the edges between $V_1$ and $V_2$. 

\begin{figure}[htbp]
\centerline{\includegraphics[width=0.2\textwidth]{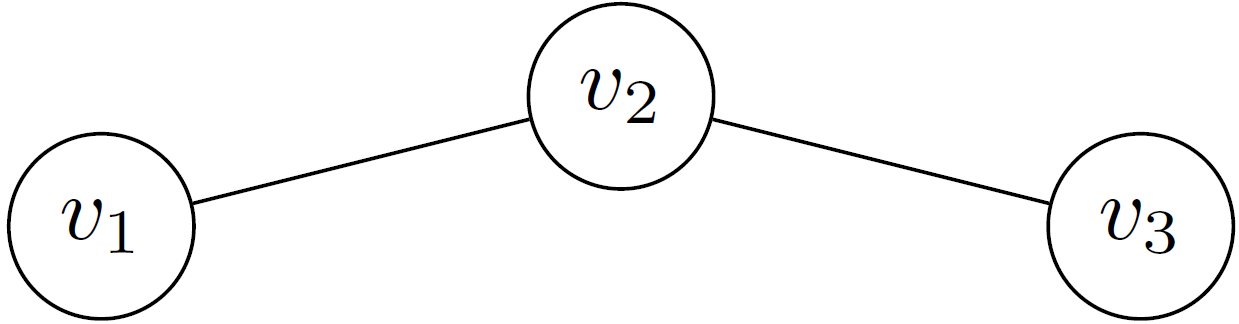}}
\caption{Example graph.}
\label{fig1}
\end{figure}

\paragraph*{Example}Consider an undirected unweighted graph $G$ that contains three vertices $\{v_1, v_2, v_3\}$ and two edges $\{(v_1, v_2), (v_2, v_3)\}$ as shown in Fig.~\ref{fig1}. If the cut is $V_1 = \{v_1\}$ and $V_2 = \{v_2, v_3\}$, or $V_1 = \{v_1, v_2\}$ and $V_2 = \{v_3\}$, the size of the cut is 1. If it is $V_1 = \{v_1, v_3\}$ and $V_2 = \{v_2\}$, then it is 2, which is also the maximum cut size for this graph.

In what follows, we assume that $X = \{x_n \cdots x_1 | x_d \in \{0,1\}, 1\leq d \leq n \}$ is a set of $2^n$ possible cuts. We further assume that $x^0$ indicates that $x=0$, while $x^1$ indicates that $x=1$. With this, each element in $X$ is $n$ bits long and represents one of the $2^n$ possible partitions of $n$ vertices into two disjoint subsets $V_1$ and $V_2$. Furthermore, if an $x_d = 1$ in $x_n \cdots x_1 \in X$ then this indicates that the d-th vertex in graph $G$ is in $V_1$. For the same partition, if $x_d = 0$, then this indicates that the d-th vertex is in $V_2$. 

The fact that an edge $(x_k,x_p) \in V_1 \times V_2$ can be verified by formula \eqref{eq1}, while the fact that an edge is not shared between $V_1$ and $V_2$ can be verified by formula \eqref{eq2}. 
\begin{equation}\label{eq1}
    f(x_k, x_p) = (x_k \wedge \overline{x_p})\vee (\overline{x_k} \wedge x_p)
\end{equation}
\begin{equation}\label{eq2}
    g(x_k, x_p) = (\overline{x_k} \wedge \overline{x_p})\vee (x_k \wedge x_p)
\end{equation}
Here, $\wedge$ stands for the logic AND, and $\vee$ stands for the logic OR operation.

\section{Related works and motivation}
Quantum computing promises to solve certain hard problems more efficiently than classical algorithms. This holds especially for the case when the input size is too large for classical algorithms to process. Some of the best known quantum algorithms that offer such a speedup are quantum integer factorization \cite{s1} which runs exponentially faster than any known classical algorithm, and quantum search algorithm \cite{g1} that offers a generic square-root speedup over classical algorithms. It has been shown \cite{b1} that classical algorithms would require $\Omega(2^n)$ queries where Grover's algorithm only requires $O(2^{n/2})$. Shor's algorithm is problem-specific. On the other hand, Grover's algorithm finds a wide range of applications as a subroutine in quantum algorithms, such as in \cite{c1,c2,w1,w2,w3}. 

While the opposite problem of finding a minimum cut in a graph is efficiently solvable by the Ford–Fulkerson algorithm \cite{f1}, the maximum cut problem is known to be NP-hard. This means that there are no known polynomial classical algorithms for the problem in general graphs. The max-cut problem in planar graphs can be however solved in polynomial time \cite{h1}. As planar graphs constitute only a small subset of the graph family, it is vital to research on finding ways to solve the maximum cut problem efficiently. 

There exist classical approximation algorithms, the best of which runs in polynomial time and has an approximation ratio of $\approx 0.878$ \cite{g2}. In contrast to that, the algorithm presented in the present work is exact.

\section{Algorithm for the MaxCut problem}

\subsection{Computing number of edges in a cut}
In order to compute the number of edges in each cut, we introduce auxiliary Boolean variables $z_{i+1,j}$ and $z_{i+1,j+1}$, $1 \leq i \leq m, 0 \leq j \leq i$. $z_{i+1,j+1}$ stores the number of edges in a cut after determining the influence of bits $(x_k,x_p)$ encoding the $(i+1)$-th edge $(v_k,v_p)$ on the number of edges (this corresponds to the number of 1s). Hence, $z_{i+1,j+1}=1$ indicates that there are $j+1$ edges in the cut. Likewise, $z_{i+1,j}$ stores the number of edges in a cut after determining the influence of bits $(x_k,x_p)$ encoding the $(i+1)$-th edge $(v_k,v_p)$ on the number of edges. $z_{i+1,j}=1$ indicates that there are $j$ edges in the cut.

Computing the number of edges in a cut is straightforward. To that end, an edge must satisfy the Boolean formula in \eqref{eq1} and the cut must have currently $j$ edges. If both conditions are met, i.e. if the formula 
\begin{equation}
    [(x_k \wedge \overline{x_p}) \vee (\overline{x_k} \wedge x_p)] \wedge z_{i,j}
\end{equation} 
is satisfied, the number of edges in a cut is increased by 1. 

Similarly, if an edge satisfies the following  formula, the number of edges remains intact:
\begin{equation}
    [(\overline{x_k} \wedge \overline{x_p}) \vee (x_k \wedge x_p)] \wedge z_{i,j}
\end{equation}

\begin{figure}[htbp]
\centerline{\includegraphics[width=0.5\textwidth]{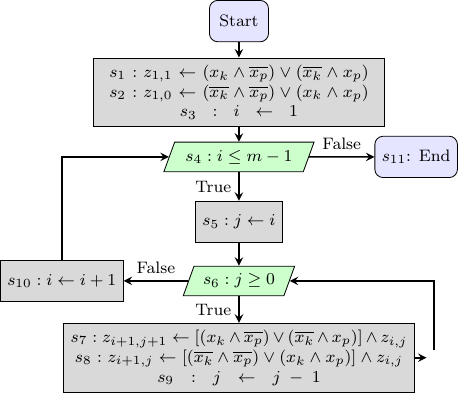}}
\caption{Flowchart for calculating the number of edges in a cut.}
\label{fig2}
\end{figure}

A flowchart on the calculation of the number of edges is given in Fig.~\ref{fig2}. 

\subsection{Quantum circuits for computing the number of edges}
Let the initial state of our quantum system be given as $\ket{\phi_0}=\bigotimes_{d=n}^1 \ket{x_d^0}$, indicating that all qubits are initialized to 0. We apply a Hadamard gate to each of the $n$ qubits, thereby obtaining the state:
\begin{equation}
    \ket{\phi_1} = H^{\otimes n} \ket{\phi_0} = \frac{\bigotimes_{d=n}^1 (\ket{x_d^0}+\ket{x_d^1})}{\sqrt{2^n}} = \sum_{x=0}^{2^n -1} \frac{\ket{x}}{\sqrt{2^n}}
\end{equation}
The state $\ket{x_n^0 \cdots x_1^0}$ in $\ket{\phi_1}$ encodes the partition of the $n$ vertices into subsets $V_1=\emptyset$ and $V_2=\{v_n,\cdots,v_1\}$. The state $\ket{x_n^0 \cdots x_2^0x_1^1}$ encodes the partition into $V_1=\{v_1\}$ and $V_2=\{v_n,\cdots ,v_2\}$, etc.

\subsection{Calculating cut membership for edges}
In order to determine to which cut an edge belongs, we introduce auxiliary qubits $\ket{r_{j,s}}, 1 \leq j \leq m(m+1)/2, 1 \leq s \leq 4$ that will store intermediate evaluation results for $x_k \wedge \overline{x_p}$, $\overline{x_k} \wedge x_p$, $\overline{x_k} \wedge \overline{x_p}$, and $x_k \wedge x_p$. Further auxiliary qubits $\ket{s_{j,1}}$, $\ket{s_{j,2}}$, $1 \leq j \leq m(m+1)/2$, initialized to 1, are used to evaluate $\ket{r_{j,1}} \vee \ket{r_{j,2}}$ and $\ket{r_{j,3}} \vee \ket{r_{j,4}}$, respectively. 

\begin{figure}[htbp]
\centerline{\includegraphics[width=0.23\textwidth]{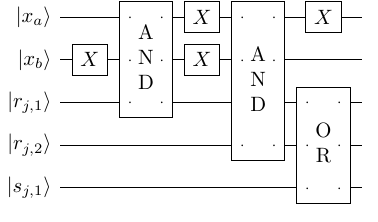}\includegraphics[width=0.23\textwidth]{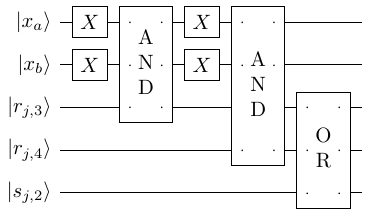}}
\caption{Left: quantum circuit EIIAC calculates formula \eqref{eq1}. Right: quantum circuit EINIAC calculates formula \eqref{eq2}.}
\label{fig3}
\end{figure}

The two circuits in Fig.~\ref{fig3} implement, respectively, \eqref{eq1} and \eqref{eq2}. 

\subsection{Calculating the number of edges in each cut}
\begin{figure}[htbp]
\centerline{\includegraphics[width=0.23\textwidth]{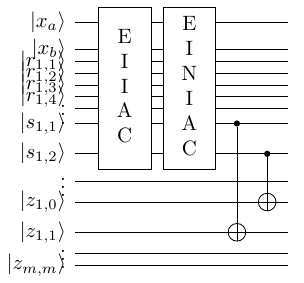}\includegraphics[width=0.23\textwidth]{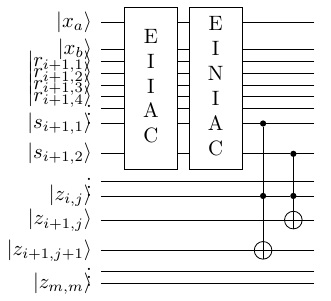}}
\caption{Left: quantum circuit CFE calculates $s_1$ and $s_2$ in Fig.~\ref{fig2}. Right: quantum circuit CSE calculates $s_7$ and $s_8$ in Fig.~\ref{fig2}.}
\label{fig4}
\end{figure}

The results of executing the operations $s_1$ and $s_2$ in Fig.~\ref{fig2} are stored in auxiliary variables $\ket{z_{1,1}}$ and $\ket{z_{1,0}}$, respectively. This fact is shown in Fig.~\ref{fig4} (left). 

Similarly, the results of executing the circuits in operations $s_7$ and $s_8$ are stored in auxiliary qubits $\ket{z_{i+1,j}}$ and$\ket{z_{i+1,j+1}}$, where $1 \leq i \leq m$ and $0 \leq j \leq i$. This fact is shown in Fig.~\ref{fig4} (right). 

\subsection{Putting the algorithm together}
The pieces of our quantum algorithm described above need to be put together and combined with Grover's algorithm for amplification of solutions. This is shown by the pseudo-code in Algorithm \ref{algorithm}. We note that $R$ in line (3) is the number of max-cuts. $R$ can be determined by the quantum counting algorithm \cite{b1}. 

The outer loop in line (1) is to be understood in such a way that if the highest value of $t$ doesn't produce a solution, then a value decreased by 1 is tested, and so on, until a solution is found. 

In Algorithm \ref{algorithm} the initial state is
\begin{equation*}
\begin{split}
     \ket{\psi_0} &= \ket{1}
    \bigotimes_{i=m}^{1}\bigotimes_{j=i}^{0}\ket{z_{i,j}^0}
    \bigotimes_{i=\frac{m(m+1)}{2}}^{1}\bigotimes_{b=2}^{1}\ket{s_{i,b}^1} \\
    &\bigotimes_{i=\frac{m(m+1)}{2}}^{1}\bigotimes_{k=4}^{1}\ket{r_{i,k}^0}
    \bigotimes_{d=n}^1\ket{x_d^0}
\end{split}
\end{equation*}

The first register ($\ket{1}$) is a standard auxilliary register used in Grover's routine for both the oracle and the diffusion operator. We will refer to it as \textit{aux}. 

After the application of Hadamard gates, the state becomes 
\begin{equation*}
\begin{split}
\ket{\psi_1} &=
     \frac{\ket{0}-\ket{1}}{\sqrt{2}}
    \frac{1}{\sqrt{2^n}}
    \bigotimes_{i=m}^{1}\bigotimes_{j=i}^{0}\ket{z_{i,j}^0}
    \bigotimes_{i=\frac{m(m+1)}{\sqrt{2}}}^{1}\bigotimes_{b=2}^{1}\ket{s_{i,b}^1}
    \\ &\bigotimes_{i=\frac{m(m+1)}{\sqrt{2}}}^{1}\bigotimes_{k=4}^{1}\ket{r_{i,k}^0}
    \bigotimes_{d=n}^1\big(\ket{x_d^0}+\ket{x_d^1}\big)
\end{split}
\end{equation*}

\begin{algorithm}
\label{algorithm}
\caption{Overview of the quantum algorithm for the maximum cut problem}\label{alg:two}
\KwData{quantum system of size $n$ qubits in state $\ket{\psi_0}$}
\KwResult{a maximum cut}
\For{$t = m$ down to $0$}{
Apply Hadamard gates to \textit{aux} and $\ket{x}$ to obtain $\ket{\psi_1}$\;
\For{$k = 1$ to $\frac{\pi}{4} \frac{\sqrt{2^n}}{R}$ }{
Apply CFE to the first edge (Fig.\ref{fig3} left)\;
\For{$i = 1$ to $m(m+1)/2$}{
  \For{$j=i$ down to $0$}{
    Apply CSE to all other edges (Fig.\ref{fig3} right)\;
  }
}
Apply CNOT gate with control qubit $\ket{z_{m,t}}$ and target qubit \textit{aux}\;
Reverse steps in lines 4-9\;
Apply diffusion operator to register $x$\;
}
\If{output shows spikes with probability $\geq 1/2$}{ break\;}
}
\end{algorithm}


\section{Complexity assessment}
As Grover's algorithm has the complexity of $O(2^{n/2})$, this is also the complexity of the present algorithm.

\section{Experimental validation}
\begin{figure}
    \centering
    \includegraphics[width=0.5\textwidth]{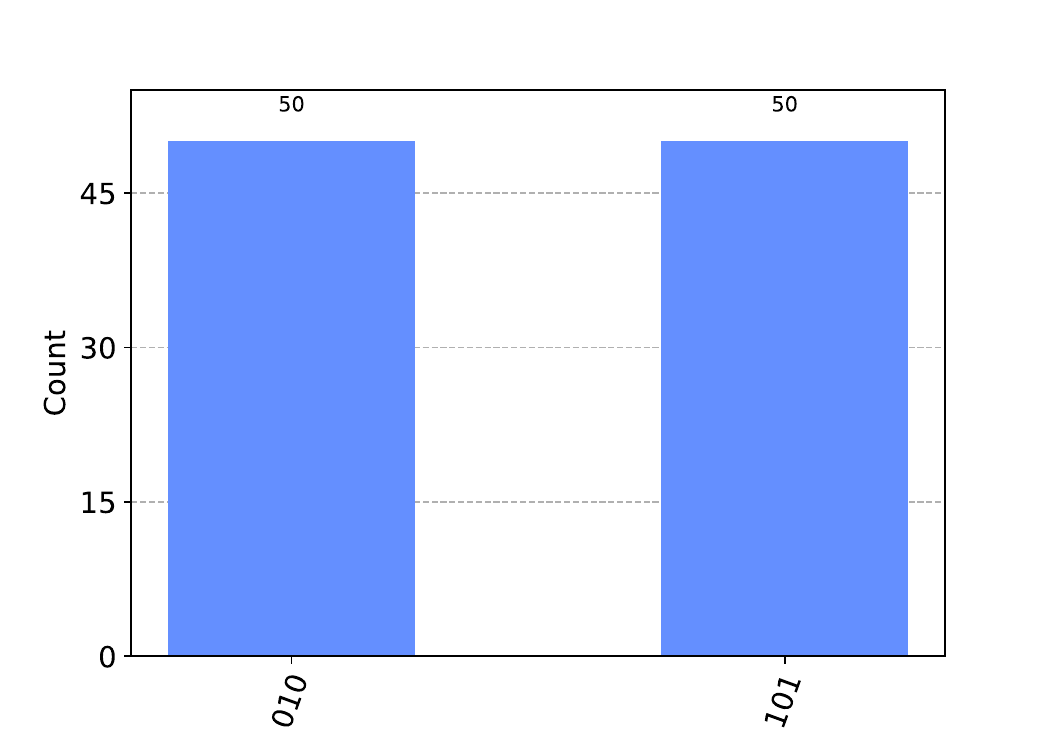}
    \caption{The maximum cut found for the example graph in Fig.\ref{fig1}. Note that both 010 and 101 indicate equivalent maximum cuts. Hence, the maximum cut for Fig.\ref{fig1} is obtained with probability 1.}
    \label{fig5}
\end{figure}

We have coded \cite{w4} and executed our algorithm on IBM Quantum qasm simulator for the example graph given in Fig.~\ref{fig1}. The statistical outcome is shown in Fig.~\ref{fig5}. The maximum cut $V_1 = \{v_1,v_3\}, V_2 = \{v_2\}$ (or \textit{vice versa}) is measured with a probability greater than 1/2.

\begin{figure}
    \centering
    \includegraphics[width=0.5\textwidth]{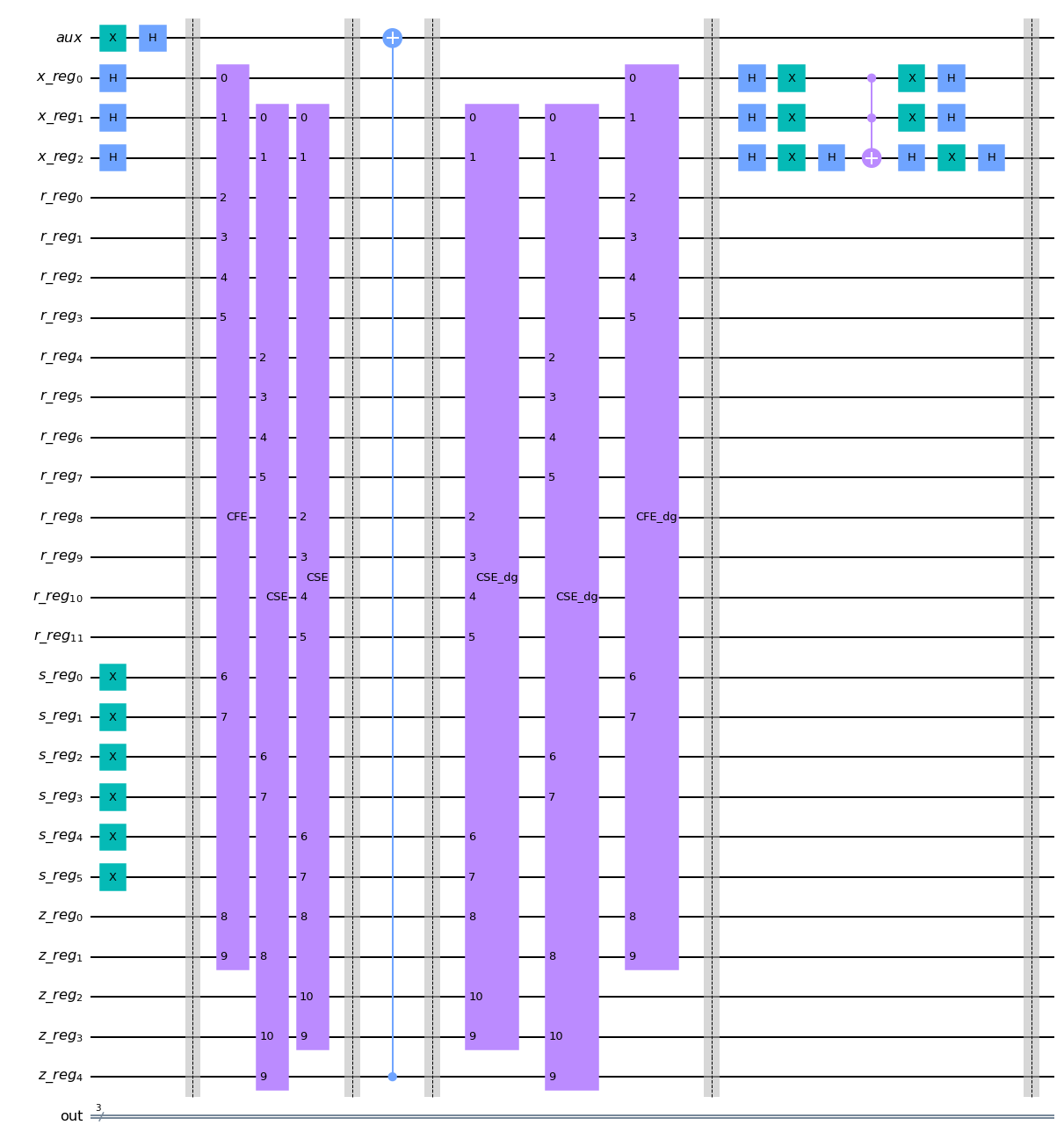}
    \caption{The circuit for the example graph in Fig.\ref{fig1}.} 
    \label{fig6}
\end{figure}

Fig. \ref{fig6} shows the circuit used to produce the outcome in Fig. \ref{fig5}. The gates applied before the first barrier belong to the initialization step. The circuit is run once only in accordance with the formula $\frac{\pi}{4}\sqrt{\frac{2^n}{R}}$, where $R=2$ since the maximum cut sets are counted twice, as indicated in Fig. \ref{fig5}. The single run consists of a block of one CFE and two CSE circuits, followed by a CNOT gate that flips the phase of the oracle qubit \textit{aux} for the case where $aux = 1$. After the CNOT gate, the CFE-CSE block needs to be uncomputed to free qubits for eventual further runs in the case such runs are specified by the formula $\frac{\pi}{4}\sqrt{\frac{2^n}{R}}$. And lastly, we have a diffusion block that amplifies the solution in each iteration/run of the Grover routine.

\section{Conclusions}
In the present paper, we have devised a quantum algorithm for the max-cut problem that offers a quadratic speedup over its classical, exact counterparts. We have further successfully executed an instance of the proposed quantum algorithm using IBM's Qiskit SDK \cite{q1}. 

\section*{Code availability}
The Python/Qiskit code for the proposed algorithm can be obtained from R. Wong's GitHub repository \url{https://github.com/renatawong/quantum-maxcut} \cite{w4}.


\vspace{12pt}
\color{red}

\end{document}